\documentclass[ %
 aip,
 amsmath,amssymb,
preprint,%
 reprint,%
]{revtex4-1}

\usepackage{graphicx}
\usepackage{dcolumn}
\usepackage{bm}

\usepackage[utf8]{inputenc}
\usepackage[T1]{fontenc}
\usepackage{mathptmx}
\usepackage{etoolbox}

\usepackage{hyperref}
\usepackage{academicons}
\usepackage{xcolor}
\usepackage{orcidlink}

\graphicspath{{./}{figures/}}
\usepackage{float}
\usepackage{amsmath}
\usepackage{amsbsy}
\usepackage{hyperref}

\newcommand{\vect}[1]{\boldsymbol{#1}}
\newcommand{\xc}{x_\mathrm{c}}
\newcommand{\nb}{n_\mathrm{b}}
\newcommand{\nd}{n_\mathrm{d}}

\newcommand{\omegacinv}{\omega_\mathrm{c}^{-1}}
\newcommand{\rhoc}{\rho_\mathrm{c}}

\newcommand{\me}{m_\mathrm{e}}

\usepackage{array}
\usepackage{multirow}
\usepackage{lineno}

\usepackage{xcolor}
\hypersetup{
    colorlinks,
    linkcolor={blue!80!white},
    citecolor={blue!80!black},
    urlcolor={blue!80!black}
}
\makeatletter
\def\@email#1#2{%
 \endgroup
 \patchcmd{\titleblock@produce}
  {\frontmatter@RRAPformat}
  {\frontmatter@RRAPformat{\produce@RRAP{*#1\href{mailto:#2}{#2}}}\frontmatter@RRAPformat}
  {}{}
}%
\makeatother
\begin{document}

\preprint{Application of mesh refinement to relativistic magnetic reconnection}

\title[Application of mesh refinement to relativistic magnetic reconnection]{Application of mesh refinement to relativistic magnetic reconnection}

\author{Revathi Jambunathan \orcidlink{0000-0001-9432-2091} }
\homepage{Corresponding author : rjambunathan@lbl.gov }
\affiliation{Lawrence Berkeley National Laboratory, 
1 Cyclotron Road, 
Berkeley, CA 94720, USA}

\author{Henry Jones \orcidlink{0000-0003-4230-5681} }
\affiliation{Lawrence Berkeley National Laboratory, 
1 Cyclotron Road, 
Berkeley, CA 94720, USA}

\author{Lizzette Corrales \orcidlink{0000-0003-0692-7547}}
\altaffiliation[Currently at ]{Cornell University, Ithaca, USA}
\affiliation{Lawrence Berkeley National Laboratory, 
1 Cyclotron Road, 
Berkeley, CA 94720, USA}

\author{Hannah Klion \orcidlink{0000-0003-2095-4293}}
\affiliation{Lawrence Berkeley National Laboratory, 
1 Cyclotron Road, 
Berkeley, CA 94720, USA}

\author{Michael E. Rowan \orcidlink{0000-0003-2406-1273}}
\affiliation{Advanced Micro Devices, Inc., 
Santa Clara, CA, USA}

\author{Andrew Myers \orcidlink{0000-0001-8427-8330}}
\affiliation{Lawrence Berkeley National Laboratory, 
1 Cyclotron Road, 
Berkeley, CA 94720, USA}

\author{Weiqun Zhang \orcidlink{0000-0001-8092-1974}}
\affiliation{Lawrence Berkeley National Laboratory, 
1 Cyclotron Road, 
Berkeley, CA 94720, USA}

\author{Jean-Luc Vay \orcidlink{0000-0002-0040-799X}}
\affiliation{Lawrence Berkeley National Laboratory, 
1 Cyclotron Road, 
Berkeley, CA 94720, USA}


\begin{abstract}
During relativistic magnetic reconnection, antiparallel magnetic fields undergo a rapid change in topology, releasing a large amount of energy in the form of non-thermal particle acceleration. This work explores the application of mesh refinement to 2D reconnection simulations to efficiently model the ineherent disparity in length-scales. We have systematically investigated the effects of mesh refinement and determined necessary modifications to the algorithm required to mitigate non-physical artifacts at the coarse-fine interface. We have used the ultrahigh-order Pseudo-Spectral Analytical Time-Domain (PSATD) Maxwell solver to analyze how its use can mitigate the numerical dispersion that occurs with the finite-difference time-domain (FDTD) (or ``Yee'') method. Absorbing layers are introduced at the coarse-fine interface to eliminate spurious effects that occur with mesh refinement. We also study how damping the electromagnetic fields and current density in the absorbing layer can help prevent the non-physical accumulation of charge and current density at the coarse-fine interface. 
Using a mesh refinement ratio of 8 for two-dimensional magnetic reconnection simulations, we obtained good agreement with the high resolution baseline simulation, using only 36\% of the macroparticles and 71\% of the node-hours needed for the baseline. The methods presented here are especially applicable to 3D systems where higher memory savings are expected than in 2D, enabling comprehensive, computationally efficient 3D reconnection studies in the future. 

\end{abstract}

\maketitle

\section{Introduction} \label{sec:intro}

Magnetic reconnection is a fundamental process where the topology of magnetic fields rapidly rearrange (they break and reconnect) converting energy stored in the stressed regions of strong magnetic fields to non-thermal particle energy. This process is often invoked to explain particle energization leading to high-energy emissions in a wide range of plasma systems. These systems include solar flares, extreme astrophysical systems such as pulsars, active galactic nuclei, gamma-ray bursts, black hole jets as well as laboratory astrophysics and even magnetic fusion devices~\citep{ji2022}. Studying the plasma kinetic effects underpinning particle energization is critical to understanding high-energy emissions from astrophysical systems. Therefore, a first-principles approach is required to capture the complex interaction of charged particles with the electromagnetic fields in these systems.

We use a fully-kinetic, electromagnetic particle-in-cell (PIC) approach~\citep{birdsall2004,arber2015,taflove2000computational} to study relativistic reconnection physics. Significant work has been done in the past two decades investigating the kinetic effects that are important to 2D relativistic reconnection in collisionless pair plasmas ~\citep{zenitani:01,zenitani:07,cerutti:12_particle_accel_crab,sironi:14,nalewajko:15,werner:16} as well as electron-ion plasmas~\citep{melzani:14}. These studies showed that the high-aspect ratio current sheets become unstable to the tearing mode instability, leading to formation of trapped plasma islands, called plasmoids, that  undergo merging and secondary reconnection during the non-linear phase. Detailed investigations on the mechanisms that drive the onset of reconnection and phases of particle energization have also been performed~\citep{guo:19,hakobyan:21,sironi:22}. The particle energy spectra due to reconnection show hard power laws that extend to high energies ~\citep{werner:16,werner:18,guo:15,hakobyan:21,petropoulou:19}.
These spectra can then be combined with radiation models to predict observational signatures of reconnection in astrophysical systems~\citep{cerutti:13, nalewajko:18}. For extreme astrophysical processes, additional quantum electrodynamics (QED) processes, such as pair-production and synchrotron radiation are important. PIC simulation studies including these effects have been reported recently ~\citep{Hakobyan2019,schoeffler2019bright,Mehlhaff2020} for 2D systems.

While substantial work has been done in 2D, similar detailed investigations of 3D systems are lacking due to the computational intensity of the PIC method. Nevertheless, a few 3D studies performed recently reveal that in addition to the tearing-mode, drift-kink instabilities dominate the evolution of the current sheet in the out-of-plane dimension~\citep{cerutti2014, guo:15,sironi:14, werner:21}, significantly complicating the picture and allowing particles to escape plasmoids. Escaped particles can re-enter reconnecting regions and become multiply-energized\citep{sironi2020}.
\citet{cerutti:14_recon_rr} investigated the dispersion relations of the tearing mode (in 2D) and the drift-kink mode (in 3D). 
Numerical investigations performed by \citet{werner:21} showed that in systems with a large guide-field (i.e., out-of-plane magnetic field component), 3D instabilities are suppressed and 2D simulations can be used as a proxy to study these systems. However, with small guide-fields, 3D effects become important, and cannot be accurately represented by 2D simulations.
More detailed investigations of 3D systems with radiation physics and QED effects have not yet been conducted, especially with large magnetization, mainly due to the computational expense of the PIC method.

Development and application of advanced numerical algorithms can improve computational efficiency and thereby enable detailed studies of 3D reconnection systems. In \citet{klion:23}, we used the ultrahigh-order pseudo-spectral analytical time-domain (PSATD) Maxwell solver and found that for 2D uniform grid simulations with same spatial resolution, it can also accurately capture reconnection similar to the widely used FDTD Yee solver. By contrast, it is not restricted by a Courant-Friendrichs-Lewy (CFL) stability criterion in theory, and our results showed excellent agreement up to CFL$=$1.6 (i.e., $c\Delta t /\Delta x=1.6$, where $c$ is the speed of light, and $\Delta t$ and $\Delta x$ are respectively the time step and the mesh size of the simulation in each direction). The simulations were performed on GPUs using the exascale-capable WarpX code, and the reconnection rate as well as particle acceleration obtained from our simulations agreed well with the results from the literature. 

Most PIC simulations reported in the literature have used a uniform grid. Note that the grid resolution for the PIC method must resolve the local skin depth in order to accurately capture the plasma kinetic effects. Magnetic reconnection involves disparate length scales wherein the plasma density is higher in the current sheet (by a factor of 5 at minimum) than in the upstream regions. In order to capture these kinetic effects, the grid resolution in the current sheet must be less than the local skin depth. But uniform resolution grids resolve even the region upstream of the current sheet, where the plasma density is lower and the corresponding skin depth is larger. Note that the PIC method also requires at-least 10s of particles per cell to obtain statistically accurate description of the non-thermal particle acceleration. Mesh refinement is therefore a natural choice to alleviate the memory requirement and improve computational efficiency. However, it has not been explored for relativistic reconnection to the best of our knowledge.

In this paper, we apply mesh refinement (MR) to relativistic magnetic reconnection using WarpX to investigate the impact of using different resolutions for the high density current sheet and low-density upstream regions. The static MR strategy we leverage was first developed by~\citep{Vay2004PML,Vay_2013} and has been previously applied to study particle accelerators and laser plasma interactions~\citep{jlvay_ieee2018,fedeli:22,lehe2022absorption}. However, it has not yet been applied to systems such as relativistic magnetic reconnection, which pose a unique set of challenges due to the high current density and large flux of particles crossing the coarse-fine interface (from upstream towards the X-points in the current sheet). While the implementation of the MR method is generalized to 3D in the code, the study of the method is performed in 2D in this work, as it enables the exploration of the key issues and their mitigation more effectively without lack of generality (i.e., the issues and mitigations identified in 2D extend readily to 3D). Previously validated, high-resolution,  uniform grid 2D simulations~\citep{klion:23} serve as the baseline to compare with the MR simulations. To study the accuracy, we compare current sheet evolution, energy conservation and conversion, and particle spectra. We also demonstrate the advantage of using a spectral method, such as PSATD, instead of the widely used FDTD method, thanks to its ultra-low numerical dispersion.

The rest of the paper is organized as follows. Sec.~\ref{sec:sim_setup} briefly describes the idealized Harris-like current sheet setup and the MR method that we used for the relativistic reconnection simulations along with our choices of numerical parameters. In Sec.~\ref{sec:MRresults}, we first present results obtained from coarsening the spatial resolution of uniform grid simulations, then from applying static MR patches surrounding the current sheet. The current sheet evolution, energy conversion, and particle energization are compared with the uniform resolution simulations, and speedups are given. Sec.~\ref{sec:MRparameters} discusses the effect of solver and parameter choices for the MR reconnection simulations that were discussed in this work. Finally, conclusions are given in Sec.~\ref{sec:conclusion} together with suggestions of improvements of the method discussed here that could be included as part of future work.

\section{Simulation Setup and mesh refinement strategy} \label{sec:sim_setup}

\subsection{Harris-Like Sheets}
\label{sec:harris}
The simulations shown in this paper are of two-dimensional, pair-plasma, relativistic magnetic reconnection, starting from Harris-like current sheets~\citep{harris:62} on a periodic domain. Since the background magnetic field changes sign at the current sheets, two sheets are needed to ensure periodicity of the magnetic field direction. This section summarizes the initial configuration; further details, including spatial profiles for all values discussed are given in our prior work, \citet{klion:23}. Unless otherwise indicated, the simulations in this paper use the same initial configuration and simulation parameters as used previously. Code and input files to replicate our simulations and results are available online. \footnote{\url{https:doi.org/10.5281/zenodo.13324091}}

The upstream, unreconnected, magnetic field is $\vect{B} = \pm B_0 \vect{\hat{z}}$. The upstream magnetic field strength $B_0$ sets the inverse upstream electron gyrofrequency, $\omegacinv \equiv \me/(eB_0)$, which is our main time scale. Here, $\me$ is the electron mass and $e$ is the elementary charge. Our base unit of length is the corresponding length scale $\rhoc = c \omegacinv$, where $c$ is the speed of light.

\begin{figure}
\begin{center}
    \includegraphics[width=1\columnwidth]{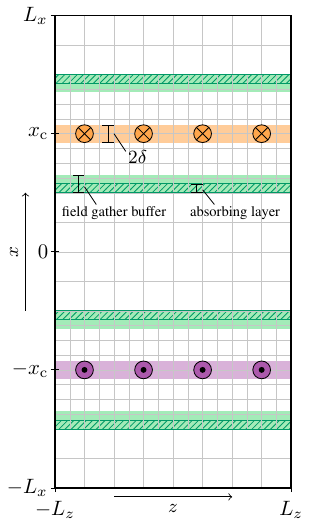}
    \caption{Initial configuration for two-dimensional relativistic magnetic reconnection with mesh refinement. The initial current sheets (orange and purple) have half-widths of $\delta$ and are located at $x = \pm x_c$. The difference in grid density demonstrates the location of the two refinement patches. Along the $x$ axis boundaries of each refinement patch, there is an absorbing layer (dark green hatched) and a larger field gather buffer (FGB) region (green highlight). The grid lines and other features on this schematic are not shown to scale.}
    \label{fig:HS_schematic}
\end{center}
\end{figure}

The 2D simulation domain extends from $-L_x$ to $L_x$ in $x$ and $-L_z$ to $L_z$ in $z$, with periodic boundary conditions in both directions. We set $L_x = 2508\,\rhoc$ and $L_z = 1254\,\rhoc$. The plasma density and bulk velocity are chosen such that they generate a $\mathrm{tanh}^2$ current density profile along the $x-$direction, centered at $x = \pm \xc \equiv \pm L_x/2$ with half-width $\delta = 12.15\,\rhoc$. The magnetic field is initialized consistently with the current sheet profile such that it satisfies Ampere's law. Expressions for these quantities are given in \citet{klion:23}. A schematic of this configuration is shown in Figure \ref{fig:HS_schematic}.

At the center of the current sheets, the magnetic field decreases to zero and therefore the magnetic pressure is also negligible. To keep the system in pressure equilibrium, the gas pressure in the current sheet must compensate for this loss of magnetic pressure. To accomplish this, the current sheet plasma is both denser and hotter than the upstream plasma. The current sheet number density per species is chosen to be $\nd = 5\nb$, where $\nb$ is the upstream number density per species. This sets the dimensionless temperature at the center of the current sheets to be $\theta_d = 1.57$. The electron-positron plasma is initialized at the start of the simulation by sampling momenta from a Maxwell-J\"{u}ttner distribution at the local temperature and with the local bulk velocity~\citep{zenitani:15}.

To initiate reconnection, a one percent sinusoidal perturbation is applied to the vector potential $\vect{A}$, which reduces the magnetic pressure at $z=0$ just above and below the current sheets. The magnetic field at initialization is the curl of this perturbed vector potential, so $\vect{\nabla} \cdot \vect{B} = 0$ at the start of the simulation. The functional form of this perturbation is given in our previous work \citep{klion:23}. The numerical methods used in this work preserve this property.

\subsection{Mesh refinement method}
\label{sec:MR_recipe}
\begin{figure}
\begin{center}
    \includegraphics[width=1\columnwidth]{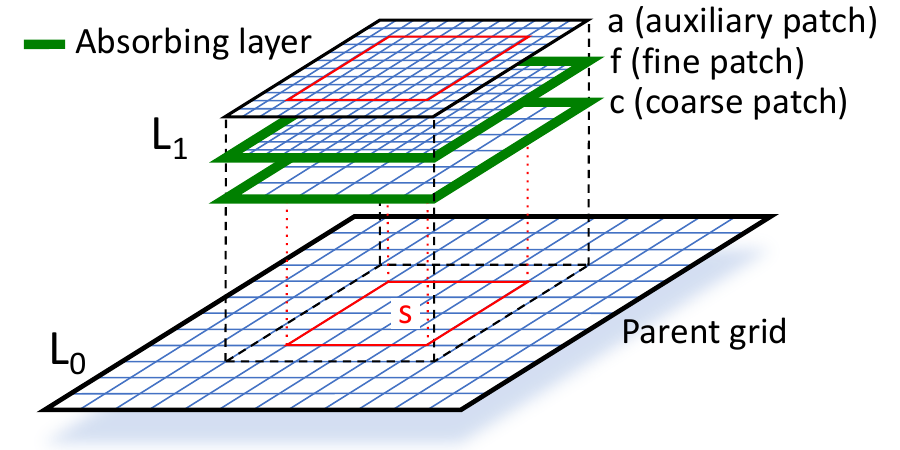}
    \caption{Schematic to illustrate the static mesh refinement algorithm, with a parent grid at Level 0, ($L_0$) and a refined region at Level 1 ($L_1$). The refined region involves three patches, namely, a fine patch, a coarse patch, and an auxiliary patch. The fine and auxiliary patches have $L_1$ resolution while the coarse patch has the same resolution as the level below it, i.e., the parent grid. Maxwell's equations are solved on the fine and coarse patches of $L_1$, and these regions are terminated by absorbing layers indicated by the green bands surrounding these patches.}
    \label{fig:MR_schematic}
\end{center}
\end{figure}

 The mesh refinement method implemented in WarpX is briefly described here; more details can be found in~\citet{Vay_2013}.
 The terminology is introduced in Fig.~\ref{fig:MR_schematic} using a simple example with one level of refinement.
The coarsest level, denoted $L_0$, is called the parent grid. The refinement region, built on top of the parent grid is referred to as level 1 ($L_1$). The overlapping area is delimited by the dotted lines on the parent grid in Fig.~\ref{fig:MR_schematic}. \textcolor{black}{The core principle of this method relies on the linearity of Maxwell's equations to separate the coarse and fine resolution solves of Maxwell's equations on every level. The total electromagnetic fields on an ``auxiliary" patch is constructed using substitution that corrects the high-resolution solutions on the fine level with the long-range effects captured on the coarse level. The particles gather the fields from these ``auxiliary" patches. In order to achieve this, every refinement region has three separate grids, referred to as fine patch, coarse patch, and auxiliary patch. The fine and auxiliary patches have the same resolution as $L_1$, and the grid resolution for the coarse patch on $L_1$ is that of the level below it, in this case the parent grid, $L_0$. Note that the parent grid does not require an auxiliary patch and the electromagnetic solution on the parent grid is the final solution for the parent grid level, $L_0$. The particles deposit their current on the fine patch of the level that corresponds to their positions, i.e., if the particle position overlaps with the refinement patch then the current is deposited on the fine patch of $L_1$. The current density on the fine patch of $L_1$ is then interpolated to the corresponding coarse patch on the same level, ($L_1$), and copied from coarse patch of $L_1$ to the parent grid $L_0$, in the region overlapping with the refinement patch (and delimited by the dotted line). Maxwell's equations are solved independently on the fine and coarse patches of $L_1$ and on the parent grid, $L_0$. The fine and coarse patches on level, $L_1$, are terminated with absorbing boundary conditions, shown by green bands in Fig.~\ref{fig:MR_schematic}.} For the parent grid, the physical boundary condition at the edge of the domain are applied. Note that in the plasma accelerator simulations that use Warp/WarpX~\citep{Vay_2013,Vay2018}, a Perfect Matching Layer (PML) was used to damp (or absorb) the electromagnetic signals leaving the fine and coarse patches of $L_1$~\citep{Vay2004PML,shapoval2019two}. This treatment at the coarse-fine boundary works well for applications like particle accelerators, where the plasma in the coarse-fine interface is nearly vacuum. However, for applications such as reconnection where the plasma is dense at the coarse-fine interface and the current density can be large, and numerical artefacts were found to build over time during reconnection with PML. In this work, we have implemented a new absorbing layer feature that is applied inside the refinement patch, as shown by the hashed region in Fig.~\ref{fig:HS_schematic} to prevent numerical artefacts that were observed with PML. The effect of using this new feature is discussed later in Sec.~\ref{sec:PML_ABC}. 

After the Maxwell solve, the full electromagnetic solution on the auxiliary patch of $L_1$ is obtained by the following substitution~\citep{Vay_2013,Vay2018}, 
\begin{equation}
    F_{1}^a(p) = F_{1}^f(p) + I^a[F_0(q) - F_{1}^c(q)](p)
\end{equation}
where $F$ is the field, $p$, and $q$ are grid points at the fine and coarse resolutions, respectively. The subscript denotes the level of the field, and the superscripts, $f$, $c$, and $a$ refer to the fine, coarse, and auxiliary patch, respectively.  $I[](p)$ is the operator that interpolates fields from the coarse resolution (q) to the grid points on the auxiliary patch of the fine resolution (p) of level $L_1$. 
For the interpolation, the solution from the coarse patch of $L_1$ is first subtracted from the solution of the underlying region from the parent grid. It is then interpolated to the higher resolution auxiliary patch, to which the solution from the fine patch, $F_{1}^f(p)$ is also added. 
Such a substitution ensures that the fine-resolution solutions (terminated at the coarse/fine boundary) are corrected to include the 
long-range interactions captured by the parent grid~\citep{Vay_2013}. 
The region $s$ delimited by a red line on the auxiliary patch of $L_1$ is offset from the coarse-fine boundary to ensure that particles do not use the fine grid solution close to the edge of the patch to avoid spurious effects that occur when the particles leave or enter the refinement patch
~\citep{Vay_2013,friedman2014}. This buffer region where particles do not gather fields close to the patch edge is referred to as the field gather buffer region, shown in green in Fig.~\ref{fig:HS_schematic}.

Simulations of relativistic magnetic reconnection with mesh refinement were initialized as follows.
The magnetic field for the Harris-sheet setup, along with the perturbation, is initialized only on the parent grid. As previously mentioned, an absorbing layer is used instead of PMLs to terminate the patches. We also deposit the current and damp it with the same damping profile in the absorbing region at the edges of the fine patch, consistent with the electromagnetic fields~\citep{lehe2022absorption}. 
The buffer gather region, starting from the edge of the absorbing layer, was chosen to be of similar physical width as the absorbing layer on the fine patch, and is indicated by the hashed region in Fig.~\ref{fig:HS_schematic}. 
The choice of these parameters for the MR simulations performed in this work are provided in Sec.~\ref{sec:MRresults}.

\section{Reconnection simulations with mesh refinement}
\label{sec:MRresults}
\begin{figure*}
\begin{center}
    \includegraphics[width=2\columnwidth]{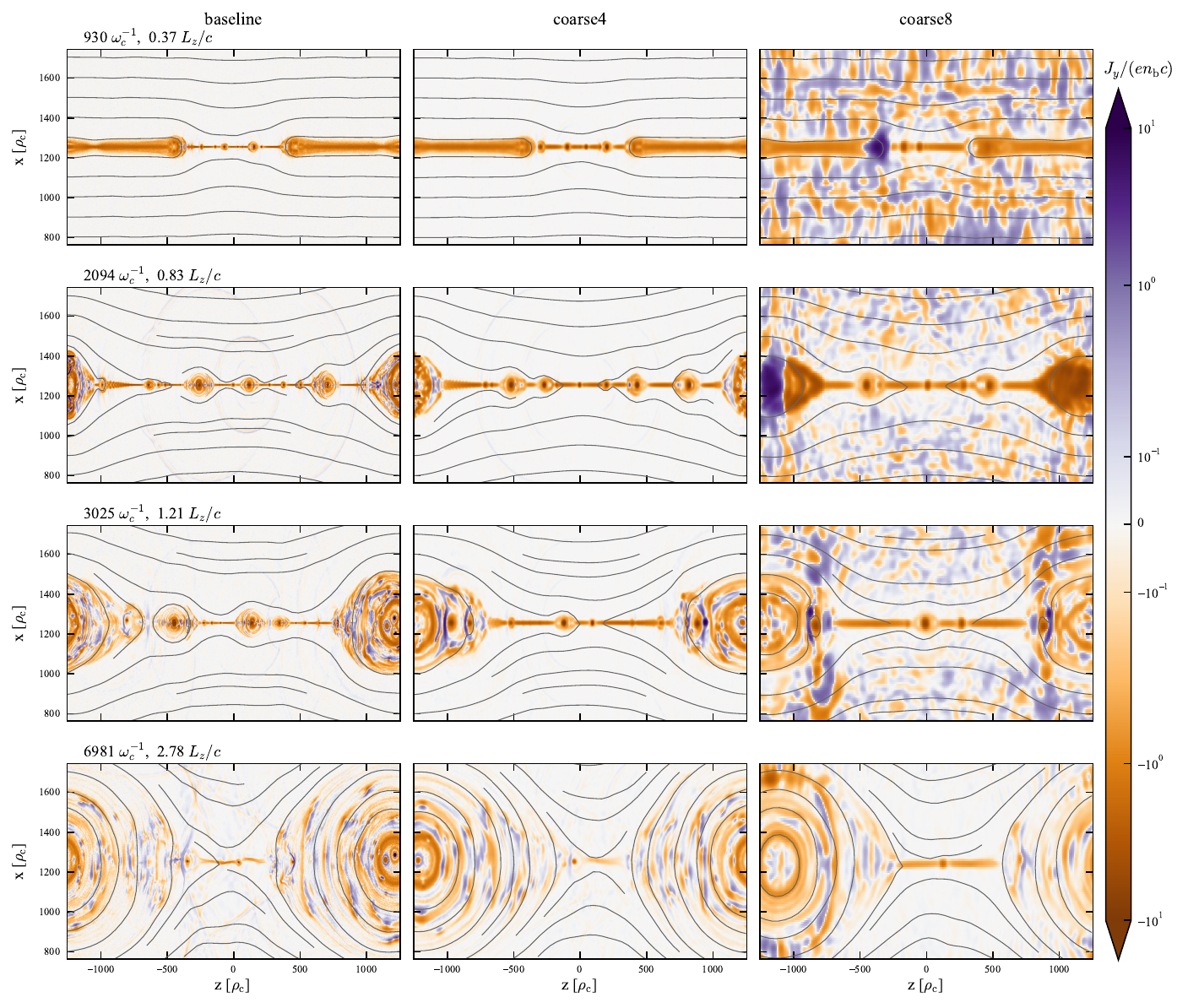}
    \caption{Comparison of temporal evolution of normalized out-of-plane current density, $j_y/(en_bc)$, obtained from three 2D uniform grid simulations with baseline, coarse4, and coarse8 resolutions. The $baseline$ and $coarse4$ simulations show qualitatively similar behavior, though the initial current sheet is somewhat underresolved in $coarse4$. The $coarse8$ simulation is strongly affected by numerical instabilities due to underresolution of the necessary plasma length scales.}
    \label{fig:jy_baseline_c4_c8}
\end{center}
\end{figure*}

\subsection{Effect of mesh refinement on current density and reconnection evolution}
\subsubsection{Effect of coarsening uniform grid resolution }


We first perform a baseline uniform grid simulation for the two-dimensional Harris-sheet set-up described in Sec.~\ref{sec:harris}, with a grid size of $4096\times2048$ cells, which has a resolution of 2 cells per current-sheet skin depth ($\Delta x = \Delta z = \lambda_e/2$). Note that the upstream Larmor radius is also resolved with nearly 1 cell (0.8 cells). Three additional uniform grid simulations are performed by coarsening the baseline simulation by factors of 2, 4, and 8, i.e., with grid size, $2048\times1024$, $1024\times512$, and $512\times256$, respectively. These cases are named \textsl{baseline}, \textsl{coarse2}, \textsl{coarse4}, \textsl{coarse8}, respectively. To differentiate the effect of the macroparticle resolution and spatial resolution of the grid, we fix the initial macroparticle resolution per unit area to be the same in these simulations, i.e., we initialize the simulations with 64, 256, 1024, and 4096 macroparticles per cell (ppc). Note that the physical plasma density at initialization is the same in all these simulations (this is achieved by the choice of macroparticle weight). 
All the simulations are performed using the PSATD Maxwell solver, with a CFL=0.95, Esirkepov deposition, and cloud-in-cell interpolation. 

A comparison of the evolution of the top current sheet obtained from the \textsl{baseline}, \textsl{coarse4}, and \textsl{coarse8} simulations are shown in Fig.~\ref{fig:jy_baseline_c4_c8}. The current sheet in the \textsl{baseline} simulation evolves similarly to the simulation performed in our previous work ~\citep{klion:23}. Soon after the current sheets are initialized, and a $1\%$  perturbation is applied, the magnetic pressure drops above and below the current sheet, causing it to collapse. The system continues to evolve, producing regions of trapped plasma called plasmoids. These plasmoids move outwards along the current sheet, and merge, forming larger plasmoids (as seen at 2094 and 3025 $\omega_c^{-1}$), finally leaving a single plasmoid at the end of reconnection (at $\sim$7000$\omega_c^{-1}$). The \textsl{coarse4} simulation qualitatively shows a similar evolution, however, due to the 4$\times$ coarser cell size, the thin current sheet is not captured as well as in the \textsl{baseline} simulation. On the other hand, the \textsl{coarse8} simulation is severely under-resolved with just 0.25 cells per current sheet skin depth, resulting in a numerical instability that appears prominently in the rightmost column of Fig.~\ref{fig:jy_baseline_c4_c8}. We note that $coarse2$ had similar evolution to $baseline$, and so have omitted it from Fig.~\ref{fig:jy_baseline_c4_c8} for brevity. 

\subsubsection{Application of mesh refinement and its effect on current density }
\begin{figure*}
\begin{center}
    \includegraphics[width=2.1\columnwidth]{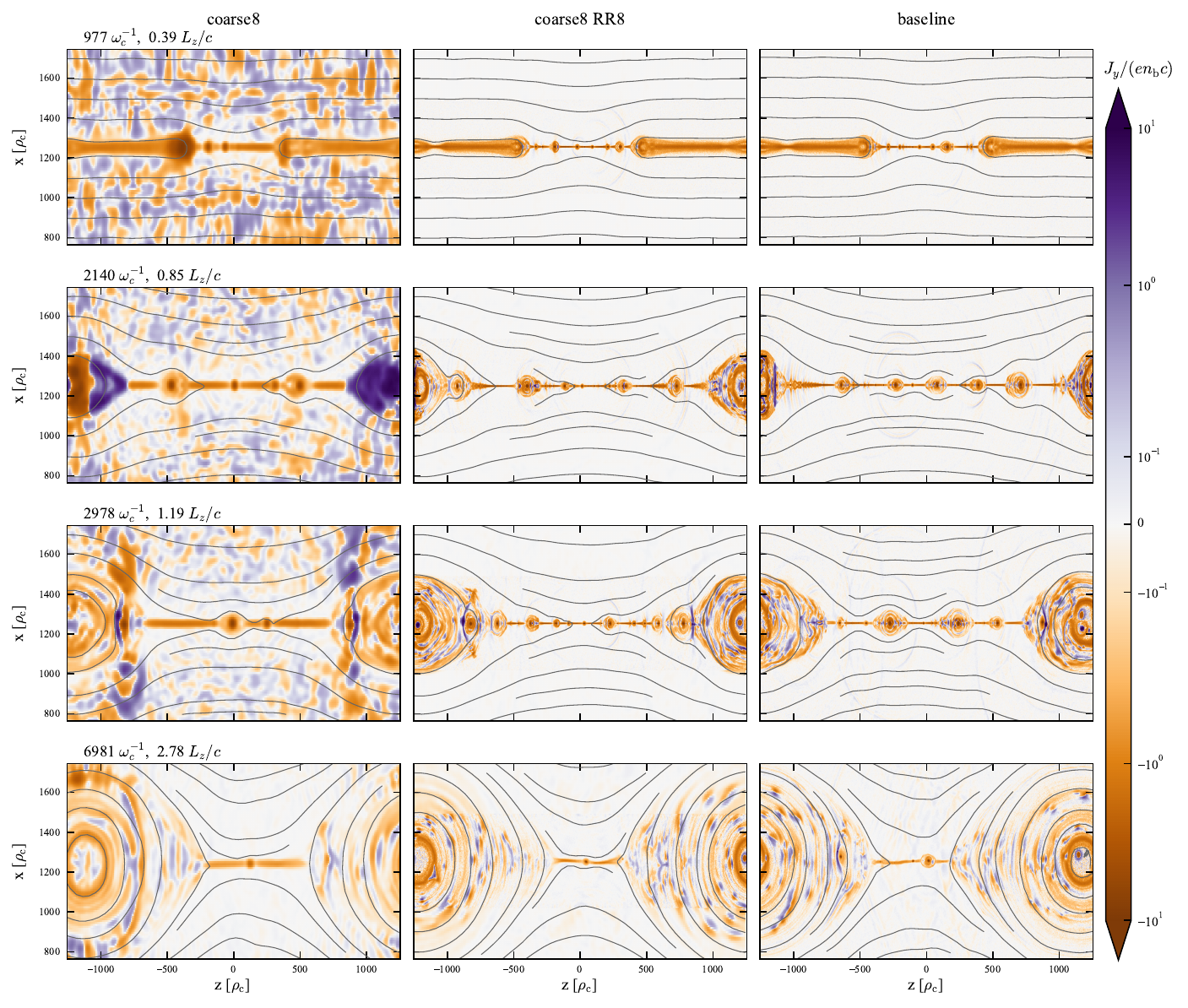}
    \caption{Comparison of temporal evolution of the normalized out-of-plane current density, $j_y/(en_bc)$, obtained from three 2D grid simulations, namely, uniform \textsl{coarse8}, mesh refinement case, \textsl{coarse8RR8}, and the \textsl{baseline} uniform grid simulation. The addition of mesh refinement greatly reduces the effects of low resolution in the $coarse8$ simulation. The qualitative evolution of $coarse8RR8$ matches that of our high-resolution $baseline$ case. Note that the fields are shown on the valid regions of $L_1$ and $L_0$ from which the particles gather electromagnetic fields.  On $L_1$, the valid region of the mesh refinement patch spans $1040<x/\rho_c<1520$ surrounding the top current sheet (shown here) and from $-1520<x/\rho_c<-1040$ surrounding the bottom current sheet (not shown here). }
    \label{fig:jy_c8_c8RR8_baseline}
\end{center}
\end{figure*}
Next, we study how the application of mesh refinement affects the results for the coarse resolution simulations.
We perform three MR simulations, with parent grids that have the same resolution as the \textsl{coarse2}, \textsl{coarse4}, and \textsl{coarse8} simulations, i.e., ($2048\times1024$), ($1024\times512$), ($512\times256$) grid-sizes. For each of these cases, static refinement patches are applied around both current sheets (as illustrated in Fig.~\ref{fig:HS_schematic}) with a refinement ratio such that the resolution of the fine patch is the same as the resolution in the \textsl{baseline} case. Refinement ratio (RR) is the ratio of the cell size on the parent grid to the refined patch. The RR for the refinement patches applied to parent grids with resolution same as \textsl{coarse2}, \textsl{coarse4}, and \textsl{coarse8} is set to RR$=2,4,$ and 8 respectively, and these MR cases are named \textsl{coarse2RR2}, \textsl{coarse4RR4}, and \textsl{coarse8RR8}, respectively. 
Note that these patches are static, therefore the refinement patch size is chosen such that it can capture reconnection physics until the end of reconnection. From our previous work, we learned that the size of largest plasmoid extends up to $\sim 800 \rho_c$. Therefore, the static mesh refinement patches are initialized to be $800 \rho_c$, resulting in 37$\%$ of the domain being refined. 
The timestep in the mesh refinement simulations is set by the CFL=0.95, based on the cell size at the finest resolution (i.e., the timestep is the same as the \textsl{baseline}) case.
 For all the mesh refinement simulations, the initial number of macroparticles per species per unit area is set to be the same as the \textsl{baseline} case, i.e., 256, 1024, and 4096 macroparticles per parent cell (64 macroparticles per fine patch cell) for the  \textsl{coarse2RR2}, \textsl{coarse4RR4}, and \textsl{coarse8RR8} cases, respectively. As mentioned previously in Sec.~\ref{sec:MR_recipe}, the refinement patch contains an absorbing boundary layer that extends into the fine patch from the coarse-fine boundary, and a field-gather buffer region, within which particles gather fields from the parent grid to avoid numerical artefacts. For all the MR simulations, the absorbing layer was set to be 10 parent grid cells wide starting from the coarse-fine interface, corresponding to 20, 40, and 80 fine-patch cells for the \textsl{coarse2RR2}, \textsl{coarse4RR4}, and \textsl{coarse8RR8}, simulations respectively. Similarly, the width of the field gather buffer region required to avoid numerical artefacts for the chosen refinement ratio are 48, 96, and 192 fine-patch cells from the coarse-fine interface (the green region in Fig.~\ref{fig:HS_schematic}) for the  \textsl{coarse2RR2}, \textsl{coarse4RR4}, and \textsl{coarse8RR8} simulations, respectively.
The numerical parameters used for the mesh refinement simulations are summarized in Tab.~\ref{tab:mrparameters}.

\begin{table}[htb]
\caption{MR simulations and parameters}
\centering 
    \begin{tabular}{|wc{7em}|wc{5em}|wc{5em}|wc{5em}|} 
    \hline
    \textbf{MR Case} & \textsl{coarse2RR2} & \textsl{coarse4RR4} & \textsl{coarse8RR8} \\
    \hline
    Parent [$N_x,N_z$] & [2048,1024] & [1024,512] & [512,256] \\
    RR                    & 2                  & 4                 & 8 \\
    Absorbing layer$^*$             & 20                 & 40                & 80 \\
    FGB$^a$ layer$^*$                   & 48                 & 96                & 192 \\
    \hline

    \end{tabular}
    \label{tab:mrparameters}\\
    \footnotesize{$^a$ Field gather buffer (FGB)}\\
    \footnotesize{$^*$ The width of the layers are set by the number of fine-patch cells}\\
\end{table}

In Fig.~\ref{fig:jy_c8_c8RR8_baseline}, we compare the evolution of the out-of-plane current density, $j_y$, obtained from \textsl{coarse8}, \textsl{coarse8RR8}, and \textsl{baseline} simulations. It can be seen that the numerical instability observed for the \textsl{coarse8} simulation is mitigated when using mesh refinement in the \textsl{coarse8RR8} case, because the mesh-refined region resolves the initial current sheet skin depth. Compared to the \textsl{baseline} simulation, it can be seen that \textsl{coarse8RR8} can capture the current sheet, the formation of plasmoids, merging of plasmoids leading to secondary reconnection (seen at $t=2978\omega_c^{-1}$), finally forming a single plasmoid at late times when reconnection has quenched. Note that, while we do not expect the evolution of plasmoids to be exactly the same as the \textsl{baseline} case, these results confirm that even with a high refinement ratio of 8, the simulations are able to capture reconnection characteristics. The evolution of current density was qualitatively similar for the \textsl{coarse2RR2} and \textsl{coarse4RR4} simulations. We chose to highlight \textsl{coarse8RR8} because the uniform grid simulation without mesh refinement, i.e., \textsl{coarse8} simulation exhibited instability, hence providing a more challenging test for mesh refinement.

\subsection{Effect of mesh refinement on energy conservation and conversion}

\begin{figure}[htb!]
\begin{center}
    \includegraphics[width=0.8\columnwidth]{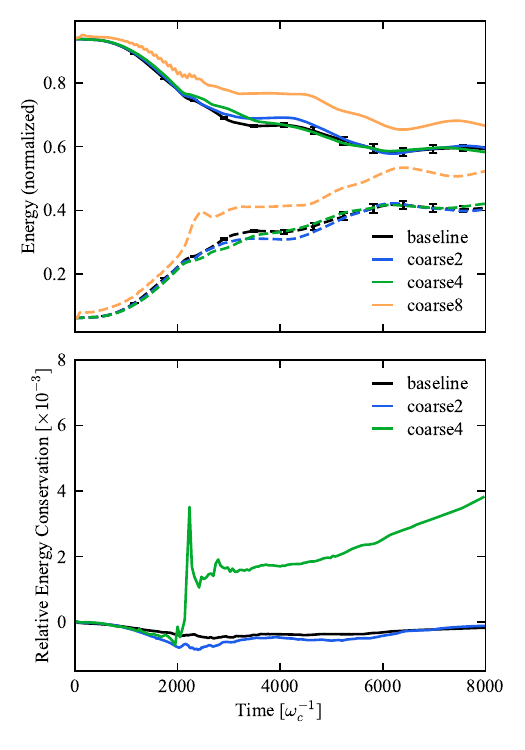}
    \caption{Comparison of energy conversion (top) and relative energy conservation (bottom) obtained from the uniform grid 2D \textsl{baseline}, \textsl{coarse2}, \textsl{coarse4}, and \textsl{coarse8} simulations. In the top panel, the magnetic field energy and particle energy are normalized by the total initial energy and shown by solid and dashed lines, respectively. The \textsl{baseline} result shown in black is averaged over five repeated simulations along with the standard deviation denoted by error bars.}
    \label{fig:energyconversion_uniform}
\end{center}
\end{figure}

A comparison of the energy transfer from magnetic field to particle kinetic energy (thermal and bulk acceleration) obtained from the \textsl{baseline}, \textsl{coarse2}, \textsl{coarse4}, and \textsl{coarse8} simulations is shown in the top panel of Fig.~\ref{fig:energyconversion_uniform}. The current sheet evolution includes a linear regime, when the current sheet breaks and forms small regions of trapped plasma, and during this time, the particle energy increases exponentially. At around $t\sim1800\omega_c^{-1}$, transition to the non-linear regime begins where plasmoids merge to form larger plasmoids also causing secondary reconnection. This continues until reconnection ceases by $t\sim7000\omega_C^{-1}$ when the magnetic field energy and particle energy reach quasi steady-state.
Since energy transfer can exhibit some small differences in the non-linear regime beginning at $t=1800\omega_c^{-1}$, five simulations were performed with the \textsl{baseline} numerical parameters. The averaged magnetic field (solid) and particle energy (dashed) obtained from these simulations are shown in black for the \textsl{baseline} case in Fig.~\ref{fig:energyconversion_uniform}, along with the standard deviation (error bars).
Energy conversion obtained from \textsl{coarse2} and \textsl{coarse4} simulations proceeds identically with the \textsl{baseline} case in the linear regime, where tearing mode instability dominates, i.e., until $t=1800\omega_c^{-1}$. Beyond this linear regime, there are some differences in energy transfer but within the standard deviation of the \textsl{baseline} for the \textsl{coarse2}, and a few standard deviations of the \textsl{baseline} for the \textsl{coarse4} case in the mid-reconnection phase ($1800\omega_c^{-1}<t<5000\omega_c^{-1}$), beyond which it is within the standard deviation until the end of reconnection. On the other hand, the \textsl{coarse8} simulation does not capture the energy transfer process accurately since the skin depth for this case is highly under-resolved, as was also observed from the current density evolution shown previously in Fig.~\ref{fig:jy_baseline_c4_c8}. 
In the bottom panel of Fig.~\ref{fig:energyconversion_uniform}, we compare the relative energy conservation and it can be seen that the \textsl{baseline} and \textsl{coarse2} energies are well-conserved and \textsl{coarse4} deviates from \textsl{baseline} when the non-linear reconnection regime begins ($t=1800\omega_c^{-1}$), however, it is still within a relative difference of $4\times10^{-3}$. The \textsl{coarse8} simulation did not conserve energy, consistent with the increase in magnetic field energy and particle energy observed in the top panel. 

\begin{figure}[htb!]
\begin{center}
    \includegraphics[width=0.8\columnwidth]{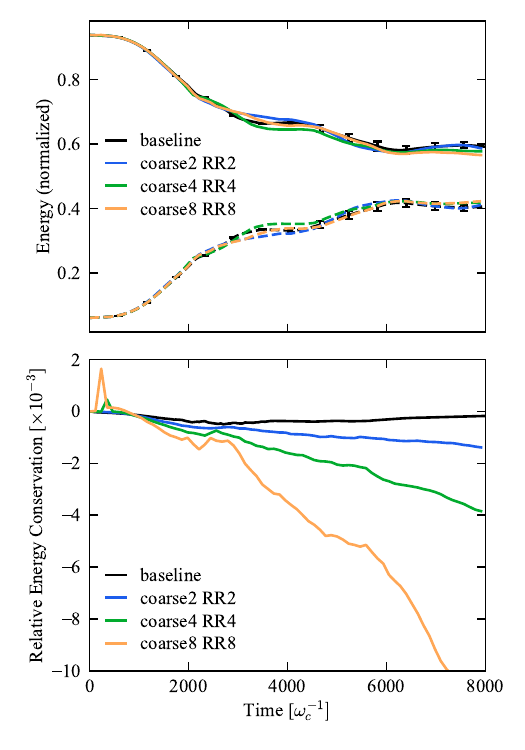}
    \caption{Comparison of energy conversion (top) and relative energy conservation (bottom) obtained from the uniform grid 2D \textsl{baseline}, \textsl{coarse2RR2}, \textsl{coarse4RR4}, and \textsl{coarse8RR8} simulations. In the top panel, the magnetic field energy and particle energy are normalized by the total initial energy and shown by solid and dashed lines, respectively. The \textsl{baseline} result shown in black is averaged over five repeated simulations along with the standard deviation denoted by error bars.}
    \label{fig:energyconversion_RR}
\end{center}
\end{figure}

In Fig.~\ref{fig:energyconversion_RR}, we compare the energy conversion obtained from the 2D mesh refinement simulations, \textsl{coarse2RR2}, \textsl{coarse4RR4}, \textsl{coarse8RR8} with the magnetic field and particle energy evolution averaged from five \textsl{baseline} simulations. As seen from the top panel, the energy conversion proceeds identically for all the cases in the linear regime (until $t=1800\omega_c^{-1}$), and then within the standard deviation from the baseline simulation. In the bottom panel we compare energy conservation (relative energy with respect to initial total energy) from the MR simulations and the baseline. By construction, the MR method is not energy-conserving, because we damp the electromagnetic fields and the current density in the absorbing layer adjacent to the coarse-fine interface in the fine patch (i.e., Level 1 grid). Even then, the energy is conserved within 1\% for the \textsl{coarse8RR8} simulation with the largest refinement ratio of 8. We investigated the initial bump at around $t=200\omega_c^{-1}$ observed for the \textsl{coarse8RR8}, and \textsl{coarse4RR4} cases, and found that it is caused by a very small signal that crosses the edges of the field-gather buffer region in the fine patch. But later on, this small bump in the signal does not affect the reconnection physics or energy conversion processes as apparent from the time evolution of energy transfer from the top panel.

\subsection{Effect of grid resolution and mesh refinement on particle acceleration}
\begin{figure}[htb!]
\begin{center}
    \includegraphics[width=\columnwidth]{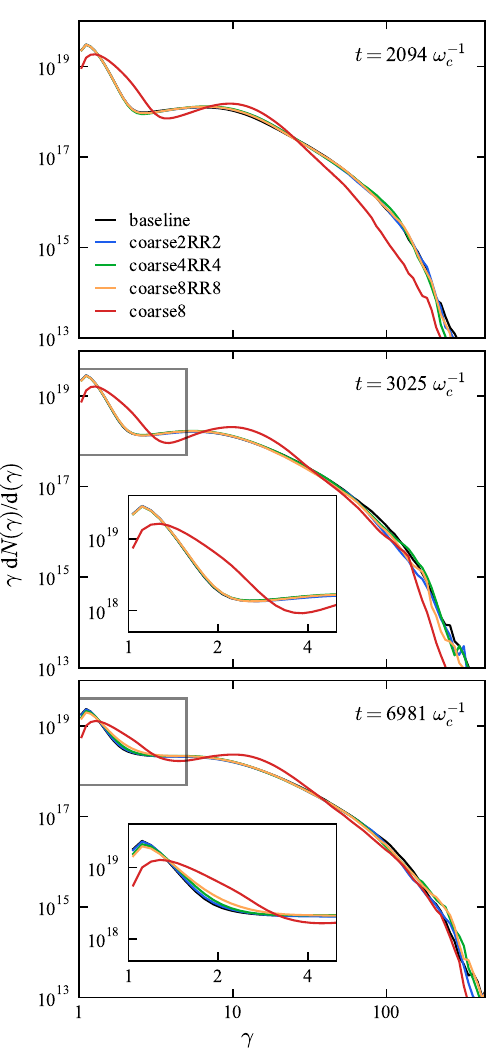}
    \caption{Comparison of time evolution of particle spectra for the \textsl{baseline} (solid black), and MR simulations, \textsl{coarse2RR2}, \textsl{coarse4RR4}, and \textsl{coarse8RR8} simulations shown by dotted lines. The solid red line is obtained from the uniform grid \textsl{coarse8} simulation.}
    \label{fig:spectra_MR_baseline_c8}
\end{center}
\end{figure}

The evolution of particle acceleration from the mesh refinement simulations, \textsl{coarse2RR2}, \textsl{coarse4RR4}, and \textsl{coarse8RR8} is compared with the \textsl{baseline} and \textsl{coarse8} uniform grid simulations, in Fig.~\ref{fig:spectra_MR_baseline_c8}. Similar to our previous work~\citep{klion:23}, the highest particle $\gamma$ at the start of reconnection is 30, and at the end of reconnection, $t=7000\omega^{-1}$, the highest particle $\gamma$ for our baseline simulation increased by an order of magnitude to 500. Majority of the particles have $\gamma \leq \sigma$, where, $\sigma=30$ is the magnetization used for our 2D relativistic reconnection simulations, which is consistent with previous results in the literature~\citep{guo:15,werner:16,werner:18}. It can be seen that compared to the \textsl{baseline} simulation, the particle spectra obtained from the \textsl{coarse8} uniform grid simulation is subject to numerical heating and does not exhibit the expected power law. This is consistent with the energy-increase observed in Fig.~\ref{fig:energyconversion_uniform} and inability to capture current sheet in Fig.~\ref{fig:jy_baseline_c4_c8}. However, for the parent grid with the same resolution, when including a mesh refinement patch as in \textsl{coarse8RR8} simulation, the particle spectra and thus particle acceleration is captured accurately. Note that \textsl{coarse2RR2} and \textsl{coarse4RR4} also show the same quantitative evolution of particle spectra. The uniform grid \textsl{coarse2} and \textsl{coarse4} simulations are not shown for brevity, but they quantitatively agree with the \textsl{baseline} simulations.

\subsubsection*{S\MakeLowercase{imulations with fewer macroparticles}}

\begin{figure}[htb!]
\begin{center}
    \includegraphics[width=\columnwidth]{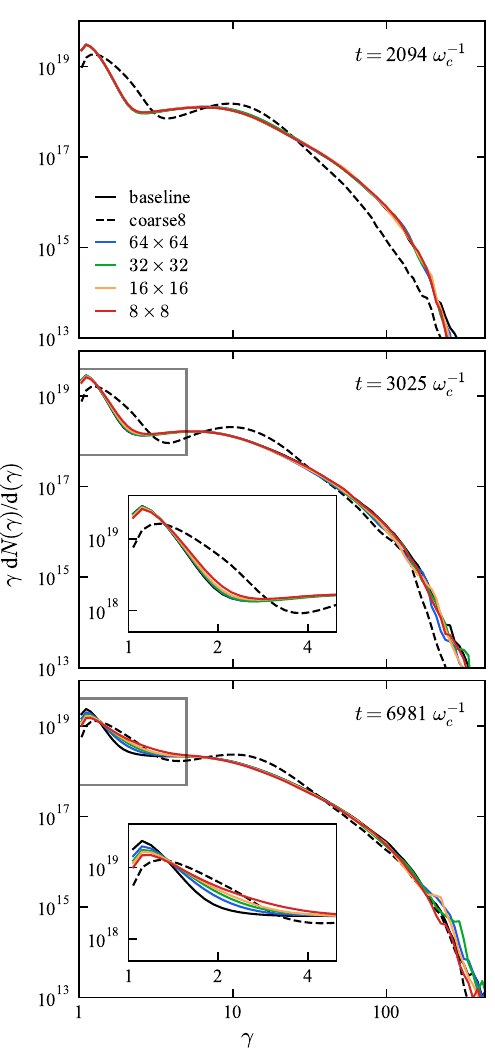}
    \caption{Comparison of time evolution of particle spectra for the \textsl{baseline} (solid black), \textsl{coarse8} (dashed black), and \textsl{coarse8RR8} simulations initialized with (64\texttimes64), (32\texttimes32), (16\texttimes16), and (8\texttimes8) macroparticles in the coarse region cells, and (8\texttimes8) macroparticles per cell in the fine-patch regions (similar to \textsl{baseline}). }
    \label{fig:spectra_particle_resolution}
\end{center}
\end{figure}
Particle-in-cell simulation runtimes are dominated by the total number of particles, especially for the relativistic reconnection simulations presented in this work. To isolate the effect of spatial resolution from macroparticle resolution, the total number of particles was kept the same for all cases presented so far (as discussed in Sec.~\ref{sec:sim_setup}). For the \textsl{baseline} and \textsl{coarse8} uniform grid simulations, there were ($8\times8$) and ($64\times64$) macroparticles per cell, respectively, at initialization, i.e., same number of macroparticles per unit area. For the \textsl{coarse8RR8} simulation also, the parent grid was initialized with ($64\times64$) particles per cell everywhere, such that the cells in the refined patch had ($8\times8$) macroparticles per species at initialization. Results are presented from three additional simulations performed with the same grid as the \textsl{coarse8RR8} case but with fewer total number of macroparticles. For these simulations, the number of initial macroparticles were sequentially reduced only in the coarse region from (64\texttimes64) to (32\texttimes32), (16\texttimes16), and (8\texttimes8) macroparticles per coarse cell per species. The number of macroparticles in the fine patch cells was maintained at ($8\times8$) for all these simulations, similar to the \textsl{baseline} and previously discussed MR simulations. 

We compare the particle acceleration obtained from these \textsl{coarse8RR8} simulations initialized with (64\texttimes64), (32\texttimes32), (16\texttimes16), and (8\texttimes8) macroparticles in the coarse regions, with the uniform grid \textsl{baseline} and \textsl{coarse8} simulations in Fig.~\ref{fig:spectra_particle_resolution}. It can be seen that at the start of the non-linear regime, at $t\sim\ 2000\omega_c^{-1}$, the particle acceleration obtained from all the \textsl{coarse8RR8} simulations with different initial macroparticle resolutions agree well with the \textsl{baseline} simulation. This agreement is also observed mid-reconnection at $t\sim 3000\omega_c^{-1}$. At the end of reconnection, the particle spectra in the \textsl{coarse8RR8} simulations with few particles agree very well with the \textsl{baseline} simulation for $\gamma > 5$. As shown in the zoomed-in inset, some minor deviation appears from the \textsl{baseline} simulation for the simulations with fewer particles. These minor differences for the low $\gamma$ region are likely due to the low macroparticle resolution in the coarse regions at the end of reconnection, since during reconnection particles from the upstream flow towards the current sheet and become trapped in plasmoids. As a result, at the end of reconnection, the upstream region has less than 5 macroparticles per cell in some regions, resulting in the low gamma region not being well-captured. \textcolor{black}{In the Appendix, we split the energy spectra into contributions from the upstream and patch regions. This further supports the statement that the difference in the spectra is caused by the under-resolution of particles in the upstream region as reconnection progresses.} Note that the lower particle resolution in the upstream region does not affect the high energy portion of the power law, which is a critical signature of reconnection, and is accurately captured even by the \textsl{coarse8RR8} simulation initialized with $8\times8$ particles per cell in the coarse region. 

\begin{table}[htb!]
\caption{Timing comparison of \textsl{baseline} and MR simulation}
\centering 
    \begin{tabular}{|wc{8.5em}|wc{4.5em}|wc{9em}|} 
    \hline
    \textbf{Case} & \textsl{Baseline} & \textsl{coarse8RR8} (64ppc$^*$)  \\
    \hline
    No. of GPU nodes & 4 & 2  \\
    No. of Particles (M)          & 1,073.7              & 413.3         \\
    Total Walltime (s)            & 560                 & 793          \\
    Total Node hours              & 0.62                 & 0.44          \\
    Performance increase                       & -                    & 1.4         \\
    \hline 
    \end{tabular}
    \label{tab:timing}

    \footnotesize{$^*$ $8\times8$ particles per cell (ppc) per species in both coarse and fine cells at initialization }\\
\end{table}

\subsection{Performance comparison with uniform grid}

Since the results obtained from the \textsl{coarse8RR8} simulation initialized with 64 particles per cell in the coarse and fine cells agree well with the \textsl{baseline} simulation, we compare the walltime and node-hours used between the uniform grid \textsl{baseline} simulation and the \textsl{coarse8RR8}. 
 Since the memory footprint for the \textsl{coarse8RR8} simulations with 8 times coarse particle resolution in the coarse patch is reduced by a factor of 4, the latter simulation fit on just two nodes of the OLCF Summit super computer while the \textsl{baseline} simulation required 4 nodes. We used 4 GPUs per node so that the grids can be equally divided between the nodes, instead of using all the 6GPUs available per Summit node. In Tab.~\ref{tab:timing}, we compare the total run-time and node hours used for the simulation. For both simulations, we did not include diagnostics or I/O, and performed the simulations up to the end of reconnection (7,000 timesteps). The \textsl{baseline} simulation with 1,073 Million particles required 560s with four Summit nodes, i.e., 0.62 node-hours. With the \textsl{coarse8RR8} simulation with 412.3 Million particles, we obtained a performance increase by a factor of 1.4 in terms of node-hours used.

\textcolor{black}{Preliminary simulations for 3D reconnection were also performed to determine the performance improvement from using a static MR for 1000 timesteps. 3D uniform grid simulations were performed with the same resolution as the 2D \textsl{coarse2} case, resulting in a ($2048\times1024\times1024$) grid, since it would be computationally intensive to have the same resolution as the 2D \textsl{baseline} case. The 3D uniform grid simulation was initialized with 8 particles per cell, resulting in 34.36 Billion macroparticles. For 1000 timesteps, the wallclock time was 119.8 s using 512 GPU nodes, i.e., 17.04 node hours. A 3D mesh refinement simulation was also performed with parent grid 4\texttimes coarser than the 3D uniform grid simulation with parent grid size of ($512\times256\times256$), similar to the resolution of the 2D \textsl{coarse8} case. Static mesh refinement patches were initialized surrounding the two current sheets, such that, 25\% of the domain was refined with a refinement ratio of 4 in each direction, and the coarse and fine cells were initialized with 8 particles per cell. The total macroparticles as a result was reduced by a factor of 3.8, requiring 8 times fewer nodes and 282~s wallclock time, i.e., 5.01 node hours. Thus with a refinement ratio of 4, and refinement patch that covered 25\% of the domain, the total performance improved by a factor of 3.37 for the 3D simulation. We also extended the MR patch to extend 37\% of the domain, and the performance increased by a factor of 1.4. Note that further investigations need to be performed for the full 3D reconnection, where, the particle distribution will become more unbalanced as reconnection proceeds. Nevertheless, these preliminary results already indicate that 3D reconnection with MR will enable higher resolutions, which might be computationally expensive or even infeasible with a uniform grid.} 

\section{Effect of mesh refinement parameters}
\label{sec:MRparameters}

\subsection{Effect of PSATD versus FDTD}
\begin{figure*}[htb!]
\begin{center}
    \includegraphics[width=2\columnwidth]{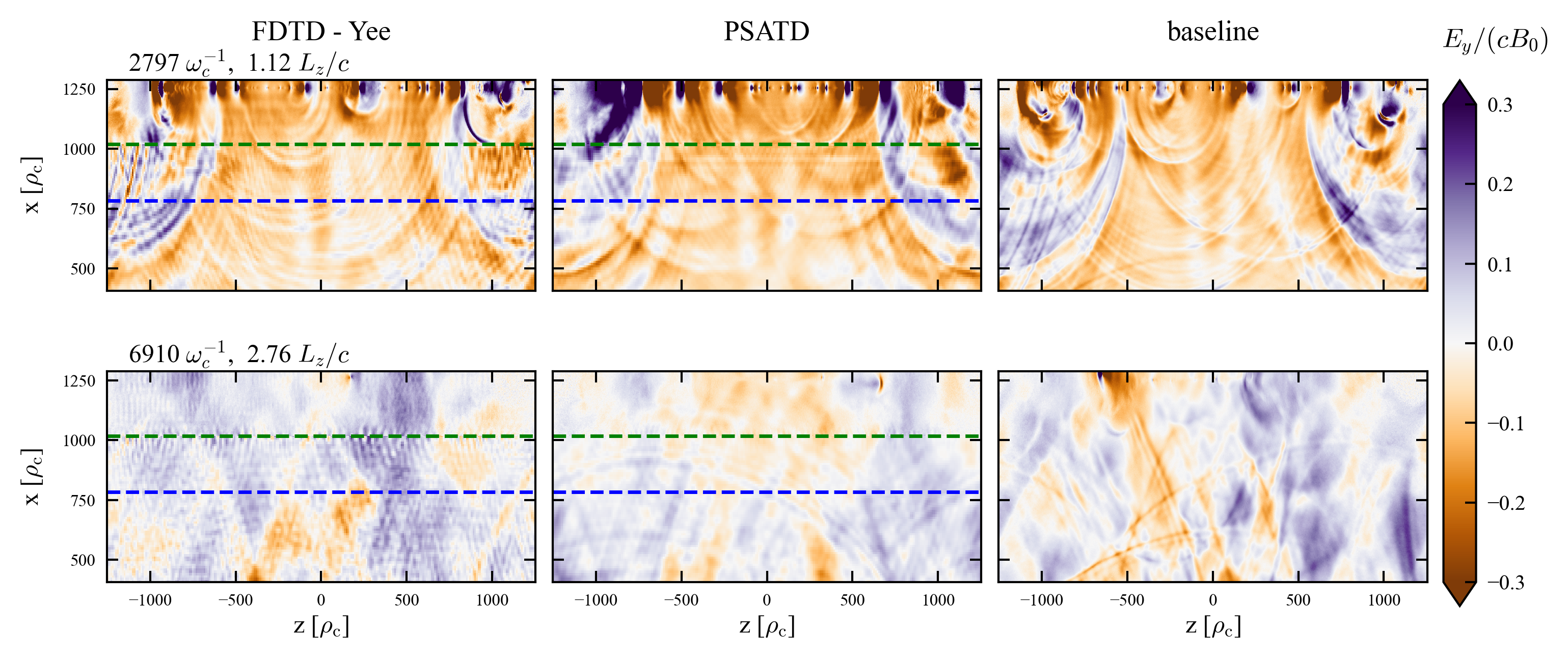}
    \caption{Comparison of spatial variation of the normalized out-of-plane electric field, $E_y/(cB_0)$ for mesh refinement simulations performed with same grid configuration as \textsl{coarse8RR8} with the Yee solver (left), the PSATD solver (middle), with the high resolution uniform grid baseline case (right). The figure is zoomed in near the coarse-fine interface, depicted by the dotted blue line, and the edge of the field-gather buffer region, depicted by the dotted green line.}
    \label{fig:yee_vs_psatd}
\end{center}
\end{figure*}

Uniform grid simulation results obtained from the widely used FDTD Yee method were compared to those obtained using the PSATD Maxwell solvers in~\citep{klion:23}, where it was found that both solvers capture the evolution of reconnection identically for the numerical parameters that were considered. In this work, we performed MR simulations for the \textsl{coarse8RR8} test case with the two solvers.  
Fig.~\ref{fig:yee_vs_psatd} shows a comparison of the normalized out-of-plane electric field for the FDTD Yee-simulation, PSATD, and \textsl{baseline} uniform grid simulation. The fine-coarse interface is shown by the dashed-blue lines for the MR simulations results. The dashed green line indicates the edge of the field gather buffer region in the fine patch, where particles in the fine-patch region between the green and blue lines gather fields from the level below (parent grid in this case). Note that the fields in these plots are from the levels that the particles gather from, and therefore, the coarse-grid result is shown in the region below the dashed-green line, since particles gather from the parent grid in those regions. At mid-reconnection, ($t\sim3000\omega_c^{-1}$), the $E_y$ solution obtained from the Yee-simulation started to develop spurious structures near the coarse and fine patch interface and by the end of reconnection, at $t\sim7000\omega_c^{-1}$, these structures are present everywhere in the domain, with larger wavelength in the coarse-grid compared to the fine-patch. A similar structure was seen previously for plasma-accelerator simulations with the Yee-scheme and the mismatch of numerical dispersion at the coarse and fine grid~\citep{jlvay_ieee2018}. Similar to previously studied numerical dispersion in the reconnection simulations we present here, the electromagnetic waves propagate at different speeds on the fine and coarse grid, which is further aggravated by the single timestep chosen to solve Maxwell's equations on every level in our simulations. Subcycling in the fine-patch with timestep ratio on the coarse and fine cells such that the corresponding CFL is the same on each grid and close to unity, was found to improve the results in previously performed accelerator simulations~\citep{jlvay_ieee2018}. 
Another solution is to use an ultrahigh-order PSATD solver.
Compared to the Yee solver, the near-dispersionless PSATD method does not develop these short-wavelength structures and the normalized electric field qualitatively compares well with the \textsl{baseline} uniform grid simulations. Thus, all the MR simulations reported in Sec.~\ref{sec:MRresults} used the PSATD solver (with order 16).


\subsection{Field damping method and parameters in absorbing layer}
\label{sec:PML_ABC}

\begin{figure*}[htb!]
\begin{center}
    \includegraphics[width=2\columnwidth]{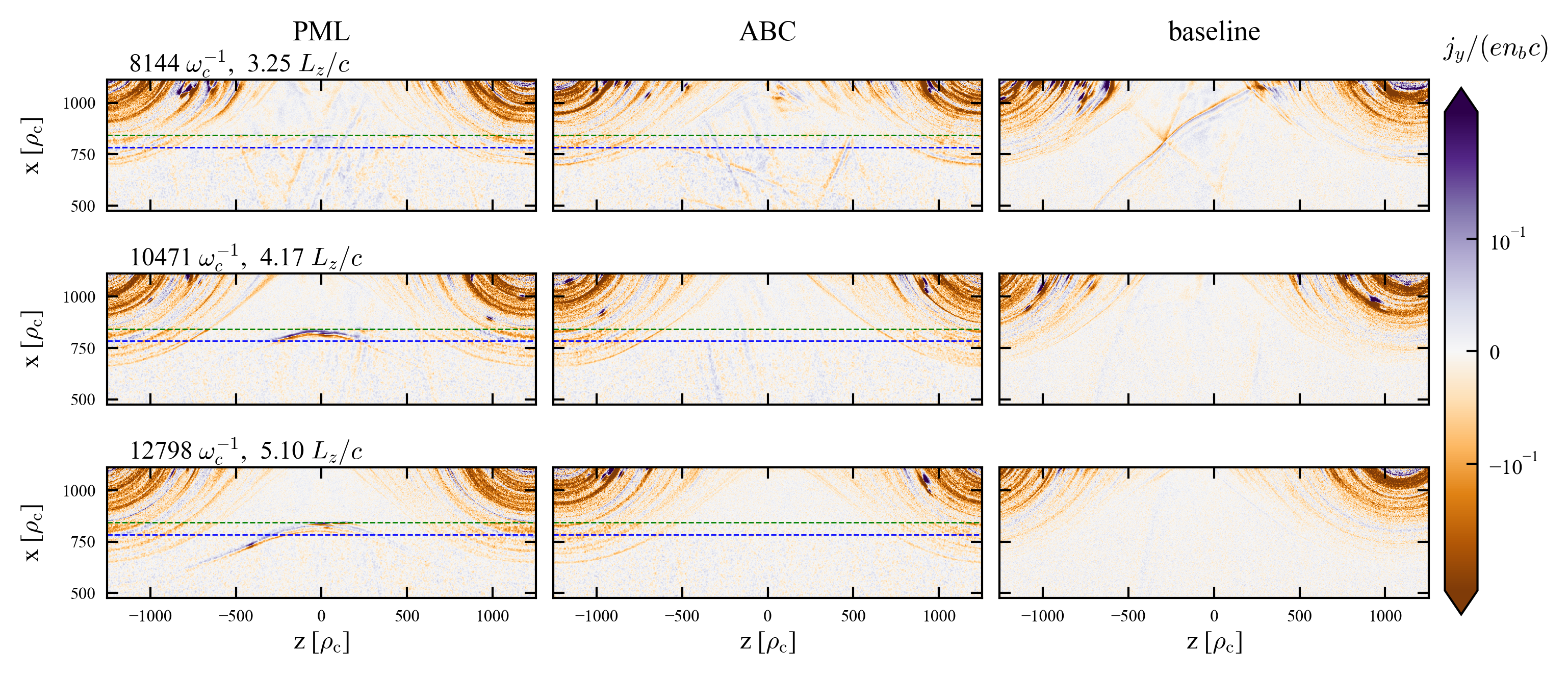}
    \caption{Comparison of spatial variation of normalized out-of-plane current density, $j_y/(en_bc)$ for mesh refinement simulations performed with \textsl{coarse2RR2} grid configuration with the PML (left), the absorbing layer (middle), and the uniform grid baseline (PSATD) case (right). The figure is zoomed-in near the coarse-fine interface, depicted by the dotted blue line, and the edge of the field-gather buffer region, depicted by the dotted green line.}
    \label{fig:jy_pml_vs_abc}
\end{center}
\end{figure*}
As mentioned previously in Sec.~\ref{sec:MR_recipe}, Maxwell's equations are solved on the fine and coarse patches of each level. At the edges of the fine patch, the electromagnetic fields are terminated by an absorbing layer. In this work, we implemented an absorbing layer to damp the fields and compared its effect to the simulations that use a Perfectly Matched Layer (PML), which is the default for mesh refinement patches in WarpX. In the PML method, the fields are split into normal and tangential components, and the normal components are damped. Since the PML method was tailored to absorb electromagnetic waves in vacuum, 
this method works well with particle accelerator simulations where typically, the plasma density near the coarse-fine interface is very small~\citep{shapoval2019two,lehe2022absorption}. However, for applications such as magnetic reconnection, where the plasma density and current density at the coarse-fine interface is high, it was found empirically that damping all the components, as is done with an absorbing layer, performs better. For damping fields in the absorbing layer, the following non-physical conductivity is used~\citep{shapoval2019two}
\begin{equation}
    \sigma_{x,i} = \sigma_{max}\bigg(\frac{i\Delta x}{\delta}\bigg)^{N_{exp}}, i = 0, ..., N_{layer}
\end{equation}
where, $\sigma_{x,i}$ is the non-physical conductivity in the $ith$ cell of the absorbing (or PML) layer, $N_{layer}$ is the number of cells in the absorbing (or PML) layer,  $\Delta x$ is the size of the cell, $N_{exp}$ sets the profile for the function (we use quadratic, $N_{exp}=2$ or cubic $N_{exp}=3$), and $\sigma_{max}=\frac{\kappa_{ds}c}{\Delta x}$ where, $c$ is the speed of light, and $\kappa_{ds}$ is the damping strength. Note that for both the PML and absorbing layer, the same conductivity profile is used to damp the fields in the absorbing layers at the edges of the refinement patches. The main difference between the two treatments, is that, in the PML, the fields are split into tangential and normal components, and only the tangential components are damped, while with the absorbing layer, all the components are damped equally.

We compare the effect of using an absorbing layer or a PML for the \textsl{coarse2RR2} case and compare with the \textsl{baseline} simulation. In addition to damping the fields, the current density from the macroparticles is also deposited in the absorbing layer and damped using the same damping profile as for the fields.
For the comparison, a cubic profile was used for the conductivity ($N_{exp}=3$), with a damping strength, $\kappa_{ds}=4$. The absorbing layer is 20 cells in the fine-patch, and the field gather buffer region is 28 cells for both simulations. The spatial variation of the out-of-plane current density, $j_y$, is shown in Fig.~\ref{fig:jy_pml_vs_abc}. We observed that until the end of reconnection, the solution with the PML method compared well with the absorbing layer and \textsl{baseline} simulation. However, when we performed these simulations for a few more light-crossing times, beyond $t>8000\omega_c^{-1}$, we observed an accumulation of non-physical current density near the PML region at the coarse-fine interface, as seen prominently at $t=10,471\omega_c^{-1}$, which continues to grow at $t=12800\omega_c^{-1}$. While the PML method did not significantly affect the solution until the end of reconnection, we decided to investigate and found that damping all the components with the absorbing layer method mitigates these numerical artefacts at the coarse-fine boundary as can be seen from the absorbing layer solution in Fig.~\ref{fig:jy_pml_vs_abc}. 
We found that the choice of the damping profile (quadratic or cubic) or the damping strength (varied from 4 to 30) does not significantly affect the solution.

\section{Conclusions and Future work}
\label{sec:conclusion}
In this work, we studied the application of static mesh refinement to first principles 2D Particle-In-Cell simulations of relativistic magnetic reconnection. Uniform grid simulations were performed first by sequentially coarsening the highest resolution uniform grid by factors of 2, 4, and 8. To distinguish the effect of macroparticle resolution and grid resolution, all the uniform grid simulations were initialized with the same total number of macroparticles (increasing macroparticles per cell by ($2\times2$), ($4\times4$), and ($8\times8$) compared to the baseline uniform grid simulation for simulations with grid coarsening factors 2, 4, and 8 respectively). The 8 times coarser grid simulation did not resolve the current sheet skin depth and therefore did not accurately model magnetic reconnection, as expected. Applying a static MR fine-patch with a refinement ratio of 8 on top of the 8\texttimes coarser parent grid led to improved resolution of the current sheet. The MR simulation was able to capture the evolution of the current sheet during reconnection, energy conversion, energy conservation, and particle spectra accurately as indicated by the excellent agreement with the uniform grid baseline simulations. The number of macroparticles was then reduced such that, at initialization, the number of macroparticles for the MR simulation with refinement ratio of 8, was $8\times8$ in the coarse and fine cells. These simulations also modeled the particle spectra and the power law accurately at all times. However, it was found that, at the end of reconnection, by which time particles from the upstream are trapped in one large plasmoid, the number of macroparticles in the coarse upstream cells are not sufficient to capture the low-energy spectra ($\gamma=2$) as accurately, and led to small deviation from the baseline solution. However, beyond $\gamma>5$, which is relevant regime for particle acceleration, the spectra compare well with the uniform resolution baseline case. 

The FDTD simulations that employed the Yee solver displayed spurious short-wavelength structures attributed to the large numerical dispersion occurring on the coarse parent grid and coarse patch. This is due to the single timestep set by a CFL$\sim$1 on the refined patch, leading to an effective CFL of 0.5 on the coarse patch. On the other hand, the ultrahigh-order PSATD solver is less susceptible to numerical dispersion, and showed good agreement with the high-resolution uniform grid baseline results. 
A new absorbing layer was introduced to reduce the numerical artefacts at the coarse-fine interface that were observed with the PML method long after reconnection quenched. 

Based on results from previous studies~\citep{klion:23}, a refinement patch was chosen that covers at-least 80\% of the largest plasmoid size expected at the end of reconnection.
This resulted in nearly 37\% of the 2D domain being refined, reducing the number of macroparticles by one half for a refinement ratio of 8. A 1.4\texttimes ~performance increase was observed in terms of node-hours used, compared to the high-resolution 2D uniform grid simulation. Manual performance optimization of the mesh refinement algorithm for magnetic reconnection, not explored here, should also provide additional performance improvements. 
Preliminary 3D uniform grid and two MR simulations were performed for 1,000 timesteps to compare the performance improvement with refined regions covering 25\% and 37.5\% of the domain. The number of macroparticles required decreased by a factor of 0.25 and 0.5, improving the performance (in terms of node hours) by a factor of 3.4 and 1.46, respectively, reducing the number of GPU nodes required by a factor of 8 and 4, respectively. Thus larger memory savings and performance increases can be expected when using mesh refinement in 3D. Detailed investigations are deferred to future work.


The MR strategies presented in this work have implications beyond the 2D relativistic reconnection application demonstrated here. The strategies presented here will also benefit non-relativistic magnetic reconnection and other high-energy systems with large disparities in length-scales and with high plasma currents crossing the coarse-fine interface. Additionally, preliminary 3D simulations show promising performance improvement, requiring fewer GPU nodes than the uniform grid counterpart. This suggests that 3D simulations with higher resolution in the current sheet are now possible due to reduced memory requirement compared to the uniform grid counterpart. This is especially significant when using radiative cooling, where cooling rates have been artificially decreased or turned off in regions where the local density becomes large, and the skin depth is  not resolved by the restrictive uniform grid \citep{hakobyan2019effects}. It will therefore open a new window to study 3D effects.

The MR strategies applied to reconnection in this work, lays the groundwork for future improvements.
The MR simulations presented here used a static mesh, and this meant using a large region for refinement even at the start of the simulation when the current sheet thickness, or region requiring refinement, is much smaller. Future work will extend this method to include adaptive refinement as the current sheet evolves to form plasmoids and as the plasmoids merge growing in size, i.e., as the region requiring refinement evolves. Similarly, the simulations used a fixed number of macroparticles with weights that were set at initialization. However, it may be more efficient to split particles when they cross from the coarse to fine region, and merge particles that transition to the coarse region. Some studies have performed particle splitting and merging \citep{dong2024dynamical}, and these will be explored in our future work. The MR algorithm presented in this work can readily be used to perform 3D simulations. While only single level MR is presented in this work, the code and methods presented here are also capable of performing multiple levels of refinement. As mentioned previously, in high-energy astrophysical systems, radiative effects are important. With the MR method presented in this work, one can resolve the local skin depth due to higher densities caused by the cooling, without having to refine the other regions. Due to the reduced memory requirement, the mesh refinement approach presented in this work will render 3D simulations more tractable as they can be performed more efficiently.

\begin{acknowledgments}

This work was partially supported by the U.S.~Department of Energy, Office of Science, Office of Advanced Scientific Computing Research, Exascale Computing Project (17-SC-20-SC) under contract DE-AC02-05CH11231, and Simulation and Analysis of Reacting Flow, FWP \#FP00011940, funded by the Applied Mathematics Program in ASCR. 
This research used resources of the Oak Ridge Leadership Computing Facility, which is a DOE Office of Science User Facility supported under Contract DE-AC05-00OR22725 with the SummitPLUS award 2023-2024.
This research also used resources of the National Energy Research Scientific Computing Center (NERSC), a U.S. Department of Energy Office of Science User Facility located at Lawrence Berkeley National Laboratory, operated under Contract No. DE-AC02-05CH11231 using NERSC award ASCR-ERCAP mp111 for 2023 and 2024. 
This research used the open-source particle-in-cell code WarpX \url{https://github.com/ECP-WarpX/WarpX}, primarily funded by the US DOE Exascale Computing Project.
Primary WarpX contributors are with LBNL, LLNL, CEA-LIDYL, SLAC, DESY, CERN, and Modern Electron.
We acknowledge all WarpX contributors and also thank Remi Lehe for the insightful discussions on mesh refinement methods used in this work.

\end{acknowledgments}



Software : 
WarpX \citep{fedeli:22,warpx_zenodo},
AMReX \citep{zhang:19,amrex_zenodo},
matplotlib \citep{matplotlib},  
numpy \citep{numpy},
scipy \citep{scipy},
yt \citep{yt}

\section*{Data Availability Statement}

The input files, submission scripts, reduced datasets, and analysis scripts to reproduce the findings of this study are openly available in \url{https://doi.org/10.5281/zenodo.13324091}

\section*{Author Declarations}
The authors have no conflicts to disclose.

\clearpage
\newpage 
\appendix

\section{Particle spectra contributions from upstream and refinement patch regions}
\label{sec:app}

\begin{figure}[htb!]
\begin{center}
    \includegraphics[width=0.7\columnwidth]{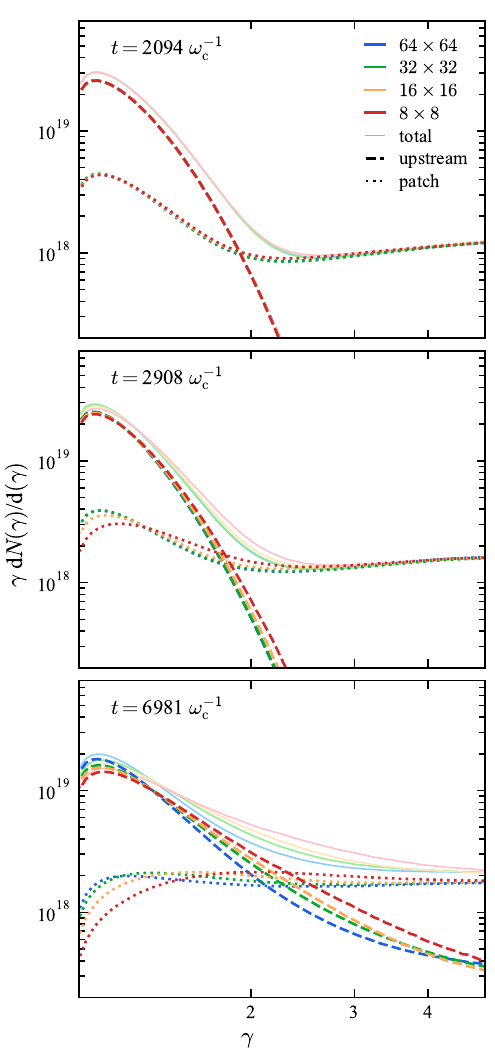}
    \caption{Time evolution of particle spectra (in the $\gamma<5$ range) obtained from the \textsl{coarse8RR8} simulations initialized with (64\texttimes64), (32\texttimes32), (16\texttimes16), and (8\texttimes8) macroparticles in the coarse region cells, and (8\texttimes8) macroparticles per cell in the fine-patch regions. The total spectra (solid) is the sum of spectra from particles in the refinement patch (dotted) and particles outside of the refinement patch called upstream (dashed). This plot focuses on the box inset region shown previously in Fig.~\ref{fig:spectra_particle_resolution} }
    \label{fig:split_spectra_particle_resolution}
\end{center}
\end{figure}

Minor differences were observed in the particle spectra (shown in Fig.~\ref{fig:spectra_particle_resolution}) obtained from \textsl{coarse8RR8} simulations initialized with (64\texttimes64), (32\texttimes32), (16\texttimes16), and (8\texttimes8) macroparticles in the coarse region cells, and (8\texttimes8) macroparticles per cell in the fine-patch regions. The energy spectra was split into contributions from particles in the refinement patch and from particles outside of the patch, called ``upstream". The spectra from these regions along with the total spectra (which is the sum of split spectra) is shown in Fig.~\ref{fig:split_spectra_particle_resolution} focusing in the $\gamma<5$ region. It can be seen that at $t\sim2000\omega_x^{-1}$, the patch (dashed), upstream (dotted), and total spectra (solid) from all simulations agree well. As time progresses, at $t=6981\omega_c^{-1}$, it can be seen that the difference in the total spectra at $\gamma=2.5$ is largely due to the particle spectra in the upstream region, where, the number of macroparticles is lower at initialization, and is further decreased during reconnection as upstream particles are pulled towards the current sheet region covered by the refinement patch. Even then, the differences are minor and as mentioned previously, the non-thermal particle acceleration is captured well by all the simulations as seen from the good agreement in the high $\gamma$ regions of the spectra in Fig.~\ref{fig:spectra_particle_resolution}.

\section*{References}

\nocite{*}
\bibliography{auto_generated_bibliography,manual_bib}{}
\inputencoding{utf8}
\end{document}